\documentclass[a4paper,11pt,final]{article}

\usepackage{fullpage}

\usepackage{amsmath}
\usepackage[notextcomp]{stix}

\usepackage{cite}
\usepackage{graphicx}
\usepackage{chessboard}
\usepackage{algorithm,algorithmic}
\usepackage[small,bf]{caption}

\usepackage[colorlinks,linkcolor=blue,citecolor=blue,urlcolor=blue]{hyperref}
\makeatletter
\renewcommand\@makefntext[1]{\leftskip=1.0em\hskip-1.0em\@makefnmark#1}
\makeatother
\newcommand{\abs}[1]{\ensuremath{|\mkern0.5mu{#1}\mkern0.5mu|}}
\newcommand{\per}{\ensuremath{\mathrm{per}\:}}
\newcommand{\PP}{\ensuremath{\mathbb{P}}}
\newcommand{\EE}{\ensuremath{\mathbb{E}}}
\renewcommand{\leq}{\leqslant}
\renewcommand{\geq}{\geqslant}

\renewcommand{\dots}{\relax\ifmmode\ldots\else$\,\ldots\,$\fi}

\begin{document}

\pagestyle{empty}
\renewcommand{\thefootnote}{\fnsymbol{footnote}}

\begin{titlepage}

\begin{center}

{\Large\textbf{Restricted permutations for the simple exclusion process \\ in discrete time over graphs}}

\vspace{6ex}

{\large\textbf{J. Ricardo G. Mendon\c{c}a}\footnote{Email: \texttt{\href{mailto:jricardo@usp.br}{\nolinkurl{jricardo@usp.br}}}}}

\vspace{2ex}

\textit{Escola de Artes, Ci\^{e}ncias e Humanidades, Universidade de S\~{a}o Paulo \\ Rua Arlindo Bettio 1000, Vila Guaraciaba, 03828-000 S\~{a}o Paulo, SP, Brazil}

\vspace{4ex}

{\large\textbf{Abstract}}

\vspace{2ex}

\parbox{32em}
{Exclusion processes became paradigmatic models of nonequilibrium interacting particle systems of wide range applicability both across the natural and the applied, social and technological sciences. Usually they are defined as a continuous-time stochastic process, but in many situations it would be desirable to have a discrete-time version of them. There is no generally applicable formalism for exclusion processes in discrete-time. In this paper we define the symmetric simple exclusion process in discrete time over graphs by means of restricted permutations over the labels of the vertices of the graphs and describe a straightforward sequential importance sampling algorithm to simulate the process. We investigate the approach to stationarity of the process over loop-augmented Bollob\'{a}s-Chung ``cycle-with-matches'' graphs. In all cases the approach is algebraic with an exponent varying between $1$ and $2$ depending on the number of matches.

\vspace{2ex}

{\noindent}\textbf{Keywords}: {Restricted permutation}~$\cdot$ {$0$-$1$ matrix}~$\cdot$ {permanent}~$\cdot$ {sequential \linebreak importance sampling}~$\cdot$ {interchange process}

\vspace{2ex}

{\noindent}\textbf{PACS 2010}: 02.50.Ga $\cdot$ 05.40.-a $\cdot$ 02.10.Ox}

\end{center}

\end{titlepage}


\pagestyle{plain}
\setcounter{footnote}{0}
\renewcommand{\thefootnote}{(\alph{footnote})~}

\section{\label{intro}Introduction}

Motivations to study exclusion processes in general and exclusion processes over graphs in particular are manifold. In physics, exclusion processes provide simple yet nontrivial models for the relaxation dynamics of a gas or fluid towards the thermodynamic equilibrium \cite{liggett85,spohn,derrida92,schutz}, together with a whole gamut of fundamental questions in statistical mechanics \cite{derrida,mallick,krauth}. They are also relevant in the modeling of biological transport at molecular and cellular levels \cite{pipkin,zia,parmeggiani}, queueing systems \cite{arita}, vehicular and pedestrian traffic \cite{rajewsky,chowdhury,traffic}, and signaling in radio and computer networks \cite{p2p,adhoc}, among others. Exclusion processes can also be viewed as generalizations of the single random walk problem on graphs and groups, an active field of investigation that has led to many developments in pure and applied probability, statistics, computer science, group theory, and harmonic analysis \cite{aldous83,aldous86,aldfill,persi,cecche,survey,saloff,mixing}, to name a few.

Exclusion processes are usually modeled as a continuous-time stochastic process, with particles attempting to jump from vertex to vertex after an exponentially distributed waiting time of parameter $1$ and succeeding if the target vertex is empty. In discrete time, mixed update schemes for exclusion processes have been proposed in the study of traffic and pedestrian dynamics using cellular automata, such as the ``shuffle updates,'' in which particles are updated exactly once per time step in a predetermined or random order within each time step \cite{kessel,wolki,rolland}. These mixed protocols avoid the difficult problem of enforcing exclusion during a synchronous update---which is exactly the problem that we address here---but are not entirely discrete-time or synchronous, since at any single update clocks tic at different (noninteger) times for different particles.

In this paper we define the symmetric simple exclusion process in discrete time over arbitrary graphs and describe a simple and efficient algorithm for its stochastic simulation. We exemplify the formalism by computing the relaxation time of the process on loop-augmented Bollob\'{a}s-Chung graphs. Research problems are mentioned in the conclusions.


\section{\label{setup}Basic setup}

Let $G=(V,E)$ be a finite connected graph of order $n$ with vertex set $V = \{1, \dots, n\}$ and edge set $E \subseteq V \times V$, and let $A$ be the adjacency matrix of $G$ with elements $a_{ij}=a_{ji}=1$ if the unordered pair $\langle i,j \rangle \in E$, usually denoted by $i \sim j$, and $a_{ij}=0$ otherwise. At our convenience, we augment $A$ by taking $a_{ii}=1$ for all $1 \leq i \leq n$ (see discussion below). To each vertex $i \in V$ we attach a random variable $\eta_{i}$ taking values in $\{0,1\}$. If $\eta_{i}=1$ we say that vertex $i$ is occupied by a particle, otherwise we say that vertex $i$ is empty. The symmetric simple exclusion process in discrete time over $G$, henceforth referred to as DTSEP($G$), is the stochastic process according to which at each integer time $ t \geq 0$ each particle on the vertices of $G$ chooses one of its neighboring vertices $j \sim i$ equally at random to jump to, with the process evolving if no vertex is targeted simultaneously by two or more particles. At any given $t$, the occupation of the vertices of $G$ is denoted by
\begin{equation}
\eta^{t} = (\eta_{1}^{t}, \dots, \eta_{n}^{t}) \in \{0,1\}^n,
\end{equation}
which we call the state of $G$. The role of the diagonal elements that we added somewhat arbitrarily to $A$ now becomes clear, for nothing in the dynamics of DTSEP($G$) precludes a particle from sojourning at its current vertex, which is equivalent to having a loop at every vertex of $G$. Moreover, such device prevents the dynamics from freezing out---think of a tree with particles stuck at the leaves (vertices of degree 1).

The DTSEP($G$) is closely related with the interchange process IP($G$), a continuous time process in which $n$ distinguishable particles hop over $G$ by means of transpositions. The IP($G$) enjoyed a revival some time ago related with a conjecture (eventually proved true) about its spectral gap \cite{jungreis,caputo,cesi,kozma}. In mathematical physics there is an analogue question of whether ferromagnetic quantum spin-$\frac{1}{2}$ Heisenberg chains display some ordering of energy levels indexed by total spin $S$ (only partially true) \cite{nachtergaele,cyclestar,ssepg}. When $G=K_{52}$, the IP($G$) describes the classic problem of shuffling a deck of cards by transpositions \cite{aldous83,aldous86,aldfill,persi,shah81}.


\section{\label{reps}Representations for the dynamics}

\begin{figure}[t]
\centering
\includegraphics[viewport=0 40 720 420, scale=0.32, clip]{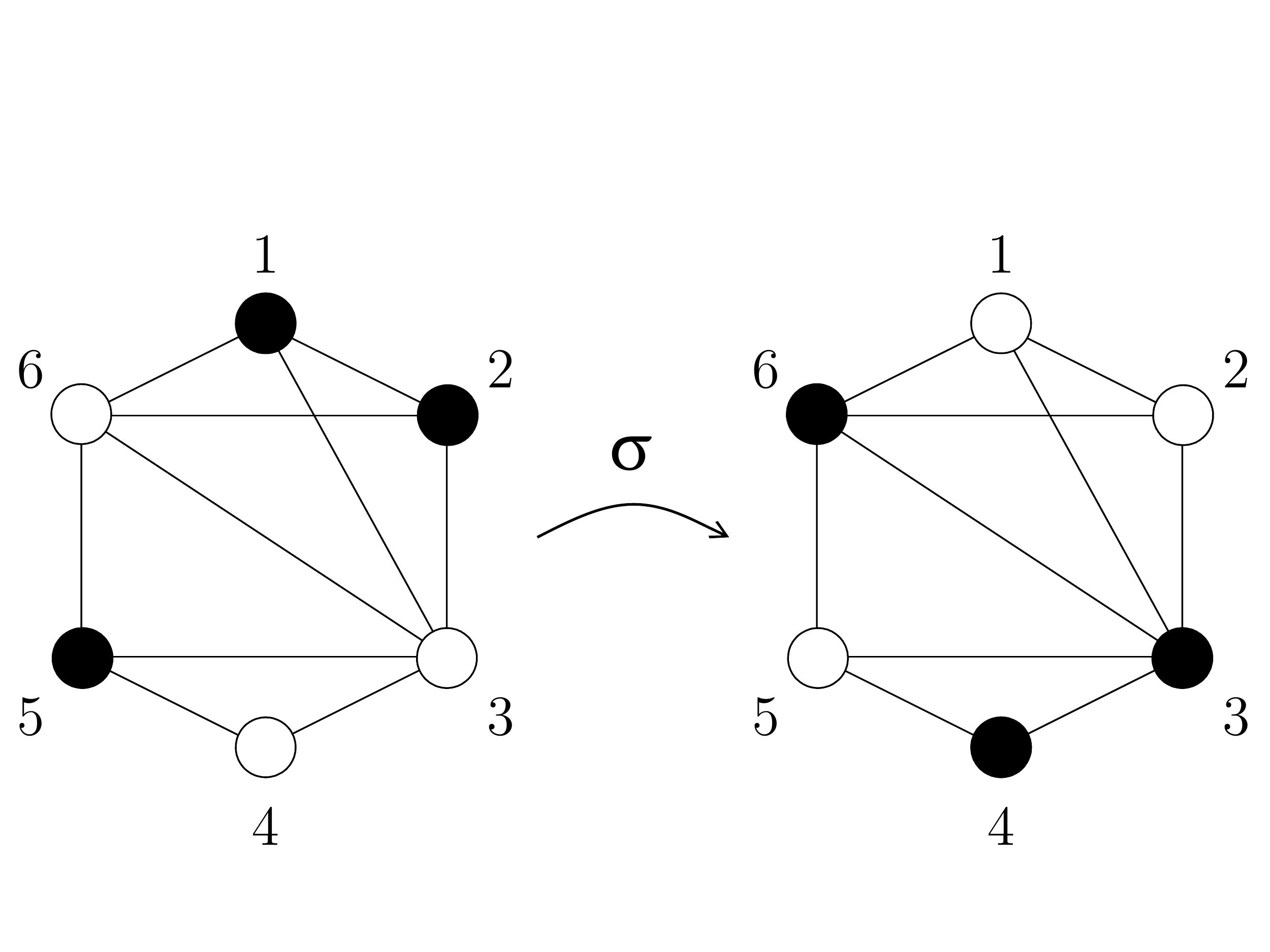}
\caption{\label{fig:dtsep}Single time step in the evolution of DTSEP($G$) over a generic graph $G$ with $k=3$ particles (black circles). The loops on the vertices are not shown for the sake of clarity. In this example the time evolution could have been driven by the permutation $\sigma=362541$.}
\end{figure}

The dynamics of DTSEP($G$) can be described by means of permutations $\sigma = \sigma(1) \cdots \sigma(n)$ in $\mathscr{S}_{n}$, the set of permutations of $n$ labels. The idea is to evolve the state of $G$ by successive applications of suitable random permutations. Permutations are convenient because they automatically conserve particles (are surjective) and enforce exclusion (are injective). Because of the restricted connectivity of $G$, however, the set of ``good'' permutations contains only permutations that take label $i$ to $\sigma(i)$ if $\sigma(i) \sim i$. This set can be characterized by
\begin{equation}
\label{restricted}
\mathscr{S}_{n}(A) =
\bigg\{\sigma \in \mathscr{S}_{n}: \prod_{i=1}^{n}a_{i\sigma(i)} = 1 \bigg\}.
\end{equation}
The number of restricted permutations in $\mathscr{S}_{n}(A)$ is given by
\begin{equation}
\label{eq:perm}
\abs{\mathscr{S}_{n}(A)} =
\sum_{\sigma \in \mathscr{S}_{n}}\, \prod_{i=1}^{n}a_{i\sigma(i)} = \per{A},
\end{equation}
i.\,e., by the permanent of $A$. Note that restricted permutations do not, in general, form a group. Pick, for example, the loop-augmented complete graph $\widetilde{K}_{4}$ (we use a tilde to discern loop-augmented graphs) and delete edge $\langle 3,4 \rangle$: then $\sigma=3412$ and $\pi=4132$ are both in $\mathscr{S}_{n}(A)$, but $\pi\sigma=3241$ is not. We note in passing that for this graph $\abs{\mathcal{S}_{n}(A)}=14$, while $\abs{\mathcal{S}_{4}}=4!=24$. We can now define the DTSEP($G$) as the stochastic process $\{\eta^{t}$, $t \geq 0\}$ that given an initial occuption state $\eta^{0}$ of $G$ evolves in discrete time according to
\begin{equation}
\label{eq:ssepg}
\eta^{t+1}_{\sigma(i)} = \eta_{i}^{t},
\end{equation}
with $\sigma$ chosen uniformly at random in $\mathscr{S}_{n}(A)$. Figure~\ref{fig:dtsep} illustrates one time step of the DTSEP($G$) on a generic graph.

Since the objects that move are the particles, all holes being indistinguishable, we can keep track of the positions of the particles instead of the occupation of the vertices. The DTSEP($G$) can thus be described in the following alternative representation. Let $\xi^{t} = (\xi^{t}_{1}, \dots, \xi^{t}_{k}) \in \{1, \dots, n\}^{k}$ be the vector of the $k \leq n$ particle positions at instant $t$. In this representation the time evolution of DTSEP($G$) is given by
\begin{equation}
\label{eq:dual}
\xi^{t+1}_{i} = \sigma(\xi^{t}_{i}),
\end{equation}
with $\sigma \in \mathscr{S}_{n}(A)$, as before. In fact, $\sigma$ now belongs to the smaller set $\mathscr{S}_{n}(A^{t})$ with $A^{t}$ the $k \times n$ matrix given by $A^{t} = (A_{\xi^{t}_{1}}, \dots, A_{\xi^{t}_{k}})^{T}$, where $A_{j}$ denotes the $j$th row of $A$. We only need to care about the full $\mathscr{S}_{n}(A)$ if $k=n$. For example, for the configurations in Figure~\ref{fig:dtsep}, $\xi^{t}=(\xi_{1}^{t},\xi_{2}^{t},\xi_{3}^{t})=(1,2,5)$ and
\begin{equation}
A^{t} = \left(\begin{array}{c}
A_{1} \\ A_{2} \\ A_{5}
\end{array}\right) =
\left(\begin{array}{cccccc}
1 & 1 & 1 & 0 & 0 & 1 \\
1 & 1 & 1 & 0 & 0 & 1 \\
0 & 0 & 1 & 1 & 1 & 1
\end{array}\right),
\end{equation}
while for $\xi^{t+1}=(\sigma(\xi_{1}^{t}),\sigma(\xi_{2}^{t}),\sigma(\xi_{3}^{t}))=(3,6,4)$ we have
\begin{equation}
A^{t+1} = \left(\begin{array}{c}
A_{3} \\ A_{6} \\ A_{4}
\end{array}\right) =
\left(\begin{array}{cccccc}
1 & 1 & 1 & 1 & 1 & 1 \\
1 & 1 & 1 & 0 & 1 & 1 \\
0 & 0 & 1 & 1 & 1 & 0
\end{array}\right).
\end{equation}
Matrix $A^{t}$ can be viewed as a $k \times n$ board of allowed particle positions at instant $t$ as well as for the next instant $t+1$, since, by definition, $a^{\,t}_{ij} = a_{\xi_{i}^{t},\,\xi_{i}^{t+1}}=1$, because $\xi_{i}^{t+1}=\sigma(\xi_{i}^{t})$ with $\sigma$ in $\mathscr{S}_{n}(A)$ or $\mathscr{S}_{n}(A^{t})$. We see that $\abs{\mathscr{S}_{n}(A^{t})}=\per{A^{t}}$ is but the number of ways $k$ indistinguishable non-taking rooks can be placed on the squares of a $k \times n$ board with the $(ij)$ square removed if $a_{ij}^{\,t}=0$ \cite{brualdi}. The ``rooks representation'' of DTSEP($G$) is illustrated in Figure~\ref{fig:rooks}. This representation makes it clear that each label $\xi_{i}$ performs an independent random walk, with exclusion ensured by the restricted permutations. The burden of DTSEP($G$) rests on $\mathscr{S}_{n}(A)$. It is also more convenient to study the dynamics of tagged particles.

\def\initial{
   \ifnum\value{ranklabel}=1 $\xi_{3}^{t}=5$ \fi
   \ifnum\value{ranklabel}=2 $\xi_{2}^{t}=2$ \fi
   \ifnum\value{ranklabel}=3 $\xi_{1}^{t}=1$ \fi
}
\def\final{
   \ifnum\value{ranklabel}=1 $\xi_{3}^{t+1}=4$ \fi
   \ifnum\value{ranklabel}=2 $\xi_{2}^{t+1}=6$ \fi
   \ifnum\value{ranklabel}=3 $\xi_{1}^{t+1}=3$ \fi
}
\begin{figure}
\centering
\setchessboard{boardfontsize=18pt,labelfontsize=10pt}
\chessboard[printarea=a1-f3, showmover=false, labelfont=\rmfamily, labelleftwidth=2em, labelleftformat=\initial, labelbottomformat=\arabic{filelabel}, inverse=false, pgfstyle=circle, padding=-1.0em, color=red, markfields={d3,e3,d2,e2,a1,b1}, setpieces={ra3,rb2,re1}]
\includegraphics[viewport=280 180 560 360, scale=0.275, clip]{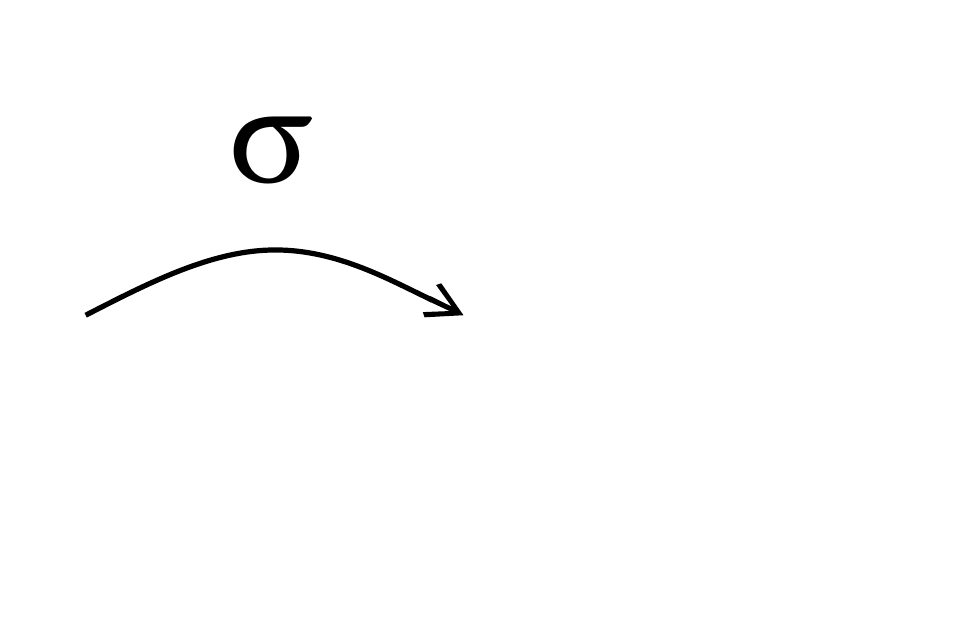}
\chessboard[printarea=a1-f3, showmover=false, labelfont=\rmfamily, labelleftwidth=2.5em, labelleftformat=\final, labelbottomformat=\arabic{filelabel}, inverse=false, pgfstyle=circle, padding=-1.0em, color=red, markfields={d2,a1,b1,f1}, setpieces={rc3,rf2,rd1}]
\caption{\label{fig:rooks}Placement of $k=3$ non-taking rooks on the boards corresponding to the particle configurations depicted in Figure~\ref{fig:dtsep}. The initial configuration $\xi^{t}=(\xi_{1}^{t},\xi_{2}^{t},\xi_{3}^{t})=(1,2,5)$ evolves through the action of $\sigma=362541$ in $\mathscr{S}_{n}(A)$ to $\xi^{t+1}=(\sigma(\xi_{1}^{t}),\sigma(\xi_{2}^{t}),\sigma(\xi_{3}^{t}))=(3,6,4)$. Marked squares indicate forbidden destinations in the next time step for the rook in the respective row.}
\end{figure}


\section{\label{simul}Stochastic simulation}

Numerically running (\ref{eq:ssepg}) or (\ref{eq:dual}) boils down to being able to sample permutations $\sigma \in \mathscr{S}_{n}(A)$ uniformly at random. A straightforward acception-rejection method would be to pick random permutations uniformly from $\mathscr{S}_{n}$ and select only those permutations for which $\prod_{i}a_{i\sigma(i)}=1$. The acceptance ratio $\abs{\mathscr{S}_{n}(A)}/\abs{\mathscr{S}_{n}}=\per{A}/n!$ of the method depends heavily on the structure of $G$, and is in general hopelessly small unless $G$ is highly dense. A much better option is to employ a sequential importance sampling (SIS) strategy. The idea behind SIS is to sample a composite object like $\sigma = \sigma(1) \cdots \sigma(n)$ by building up its parts conditioned on what has already been built according to the identity
\begin{equation}
\label{eq:p-sis}
\PP(\sigma) = \prod_{i=1}^{n} \PP(\sigma(i) \mid \sigma(1) \cdots \sigma(i-1)).
\end{equation}
The theoretical framework for SIS was given in \cite{liuchen98} and is nicely reviewed in \cite{pd-rlg-sph,cdhl}. Algorithm~\ref{alg:sis} describes a SIS strategy to sample random restricted permutations inspired by the analogous problem of estimating permanents \cite{js89,rasmussen,kuznetsov,smith,jsv04,sinclair}. Algorithm~\ref{alg:sis} can be optimized by reordering the rows and columns of $A$ in ascending order of row sums to minimize the probability of collisions between labels chosen later in the procedure with those chosen before. The extra processing pays off for graphs with vertices of widely varying degrees, as it happens, e.\,g., when $G$ is a small-world network with hubs. A careful implementation of line~\ref{alg:sis:j} (for instance, avoiding a linear search) can significantly improve its run time.

For a $0$-$1$ matrix, line~\ref{alg:sis:ri} of Algorithm~\ref{alg:sis} counts the number of images available to choose for label $i$, if any, and the probability in line~\ref{alg:sis:j} becomes the uniform distribution over the remaining images available. Note that the product of the $R_{i}$ output by Algorithm~\ref{alg:sis} provides a one-sample unbiased estimate for $\per{A}$, i.\,e., $\EE(R_{1} \cdots R_{n}) = \per{A}$ \cite{js89,rasmussen,kuznetsov,smith,jsv04,sinclair}.

\renewcommand{\thealgorithm}{S}
\begin{algorithm}[t]
\linespread{1.1}\selectfont
\caption{~Random restricted permutations by SIS}
\label{alg:sis}
\algsetup{indent=1.5em,linenosize=\small}
\begin{algorithmic}[1]
\REQUIRE $0$-$1$ matrix $A=(a_{ij})$ of order $n \geq 1$
\STATE $J \gets \{1, \dots, n\}$
\FOR {$i=1$ \TO $n$}
   \STATE Compute $R_{i}=\sum_{j\, \in\, J}a_{ij}$ \label{alg:sis:ri}
   \IF {$R_{i} \neq 0$}
      \STATE Choose $j \in J$ with probability $a_{ij}/R_{i}$ \label{alg:sis:j}
      \STATE $\sigma(i) \gets j$ \label{alg:sis:sig}
      \STATE $J \gets J \setminus \{j\}$ \label{alg:sis:J}
   \ELSE
      \STATE break
   \ENDIF
\ENDFOR
\ENSURE $\sigma(1) \cdots \sigma(n)$ is a random permutation of $1 \cdots n$ in $\mathscr{S}_{n}(A)$
\end{algorithmic}
\end{algorithm}


\section{\label{stationary}DTSEP($G$) on Bollob\'{a}s-Chung graphs}

Let $\Omega_{n,k}$ denote the set of configurations $\eta$ with $k$ particles on a single-component graph of size $n$ and let $\nu$ be the uniform measure that puts mass $\abs{\Omega_{n,k}}^{-1}={n \choose k}^{-1}$ on every $\eta$ in $\Omega_{n,k}$. Clearly, $\Omega_{n,k}$ is an invariant subspace of DTSEP($G$) and $\nu$ is stationary, since
\begin{equation}
\label{eq:stat}
\eta^{\infty} = \sum_{\eta \in \Omega_{n,k}} \nu(\eta)\eta = {n \choose k}^{-1} \mkern-4mu \sum_{1 \leq i_{1} < \cdots < i_{k}\, \leq n} \mkern-4mu (1_{i_{1}}, \dots, 1_{i_{k}})
\end{equation}
is invariant under permutations of $i_{1}$, \dots, $i_{k}$ from $\mathscr{S}_{n}(A)$, where $(1_{i_{1}}, \dots, 1_{i_{k}})$ denotes the configuration with the $k$ particles occupying vertices $i_{1}$, \dots, $i_{k}$ of $G$. Note the explicit particle-hole symmetry of the process ($0 \leftrightarrow 1: k \leftrightarrow n-k$). The occupation density of each vertex in the stationary state (\ref{eq:stat}) is $\eta^{\infty}_{i}=k/n$. On the other hand, the empirical distribution of vertex occupancy up to time $t \geq 1$ is
\begin{equation}
\label{eq:avg}
\theta^{t} = 
\frac{1}{t}(\eta^{1}+\cdots+\eta^{t}) =
\frac{t-1}{t}\theta^{t-1} + \frac{1}{t}\eta^{t},
\end{equation}
where we discard the initial $\eta^{0}$ from the average. We expect that $\theta^{t} \to \eta^{\infty}$ as $t \to \infty$. The $\chi^{2}$ distance between a realization of $\theta^{t}$ and the stationary $\eta^{\infty}$ can be calculated as
\begin{equation}
\label{eq:chisq}
\chi^{2}(\theta^{t},\eta^{\infty}) = \sum_{i=1}^{n} \frac{(\theta^{t}_{i}-\eta^{\infty}_{i})^{2}}{\eta^{\infty}_{i}}.
\end{equation}

\begin{figure}[t]
\centering
\includegraphics[viewport=5 0 490 460, scale=0.42, clip]{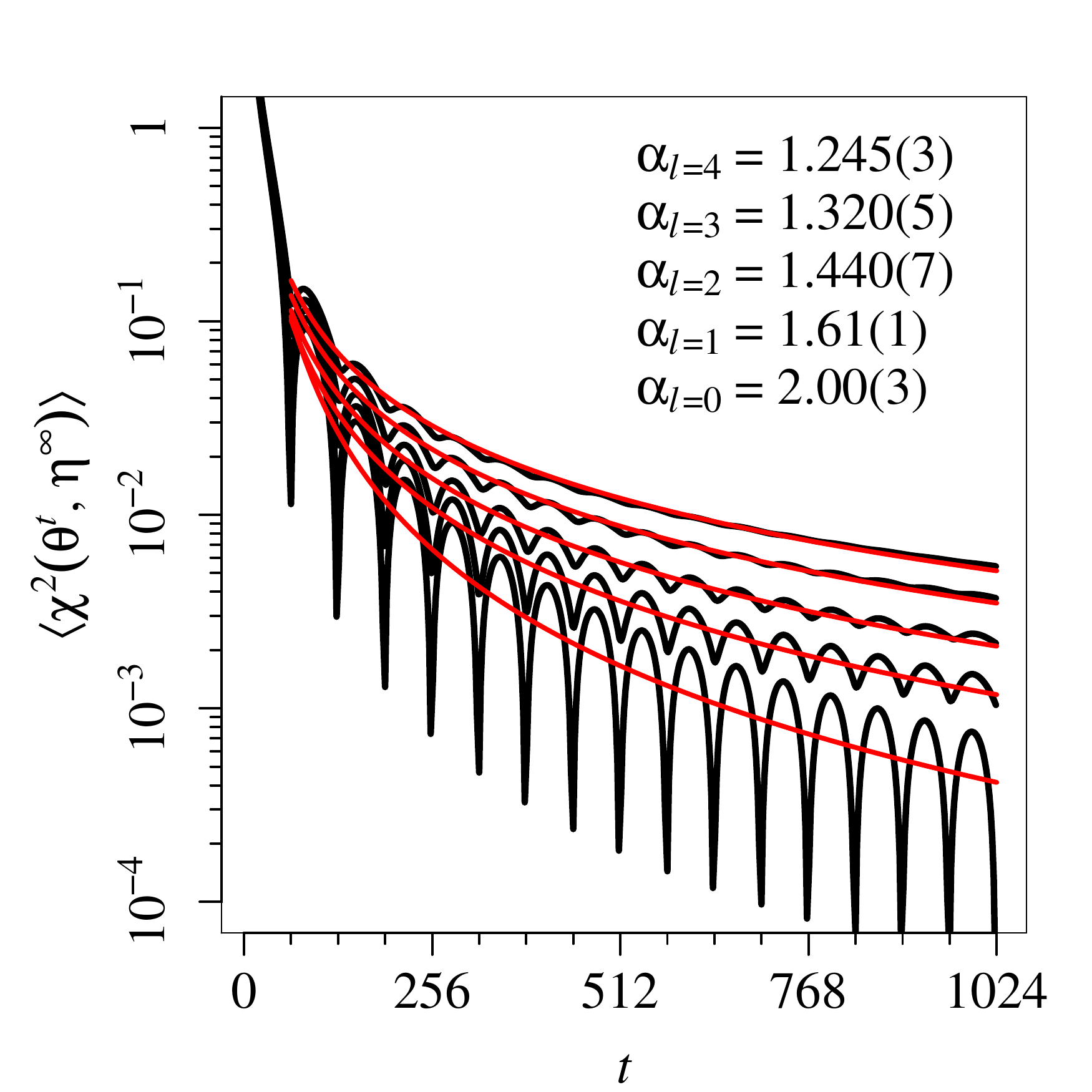}
\caption{\label{fig:relax}Averaged $\chi^{2}$ distance between the stationary and the empirical vertex occupancies on loop-augmented Bollob\'{a}s-Chung graphs with $0 \leq l \leq 4$ matches, $n=64$ vertices, and $k=16$ particles. Regression lines $\sim t^{-\alpha}$ (in red) are displayed together with the estimated $\alpha$ in each case.}
\end{figure}

We measured the speed of convergence of DTSEP($G$) to stationarity on loop-augmented Bollob\'{a}s-Chung graphs $\widetilde{C}_{n,l}$ obtained by adding $l \ll n$ (originally $l=1$) random matches (an edge $\langle i,j \rangle$ with, say, $i \leq n/2$ and $j>n/2$) to the loop-augmented cycle graph $\widetilde{C}_{n}$ \cite{belachung}. Note that $\widetilde{C}_{n,0}=\widetilde{C}_{n}$, the loop-augmented cycle graph. We fix $n=64$, $k=16$ (``quarter-filling''), and obtain $\langle\chi^{2}(\theta^{t},\eta^{\infty})\rangle$ as an average over $1000$ independent realizations of $\theta^{t}$ and, for $l \geq 1$, also over $1000$ realizations of $\widetilde{C}_{n,l}$. We found algebraic decay $\sim t^{-\alpha}$ at late times in all cases, with an exponent $1 < \alpha \lesssim 2$ depending on $l$. See Figure~\ref{fig:relax}. The ``beats'' in the $\chi^{2}$ distance at multiples of $n$ echo the cyclic structure of $\widetilde{C}_{n,l}$, which is, however, inexact for $l>0$. The $\alpha = 2.00 \pm 0.03$ for DTSEP($\widetilde{C}_{n,0}$) recalls the behavior of the simple random walk and the symmetric simple exclusion process on $\widetilde{C}_{n}$---their spectral gap closes as $n^{-2}$, and the observables approach stationarity diffusively. The discrete time version preserves that; this follows from Aldous' conjecture \cite{jungreis,caputo,cesi,kozma}. The other exponents are less immediate to understand. Simulations indicate that $\alpha \simeq 1.0$ on the loop-augmented  $\widetilde{K}_{n}$ as well as on Erd\H{o}s-R\'{e}nyi random graphs $\widetilde{G}_{n,p}$ independently of $p$ as long as the graph is simply connected. Bollob\'{a}s-Chung graphs interpolate between the two extremes given by $\widetilde{C}_{n}$ and $\widetilde{K}_{n}$. The dependence of $\alpha$ on the diameter of the graphs seems to be worth investigating in general.

\section{\label{summary}Summary and outlook}

In this paper we pursued a modest goal: to define the DTSEP($G$) and to investigate its stochastic simulation. One advantage of the setup with loop-augmented graphs (besides the fact that $\mathscr{S}_{n}(A)$ is never empty) is that one recovers the usual simple exclusion process (or, under a more general interpretation, the interchange process) over $G$ by limiting the dynamics to a single transposition per time step. The formalism applies to asymmetric exclusion processes as well, with $G$ a digraph and $A$ asymmetric. From the computational point of view, the ``rooks representation'' of DTSEP($G$) is more efficient when $k \ll n$ or $G$ is sparse, because we do not have to worry about empty vertices. This representation is also more convenient to study systems of different (or tagged) particles with different dynamics by overlaying different edge sets for different classes of particles---think of a bird flying over a $\widetilde{K}_{n}$ landscape looking after worms that crawl on a lesser graph. Discussions about reversibility, the asymmetric case, whether Algorithm~\ref{alg:sis} samples $\mathcal{S}_{n}(A)$ uniformly, comparisons with simple random walks ($k=1$), dependence of $\alpha$ on the diameter of random graphs, and related issues will be published elsewhere.


\section*{Acknowledgments}

The author thanks F\'{a}bio T. Reale (USP) for useful conversations and the S\~{a}o Paulo State Research Foundation -- FAPESP (Brazil) for partial support through grants 2015/21580-0 and 2017/22166-9.



\vspace{2ex}

\centerline{$\star$ --- $\star$ --- $\star$}


\begin{thebibliography}{00}

\bibitem{liggett85}T. M. Liggett, \textit{Interacting Particle Systems} (Springer, Berlin, 1985).

\bibitem{spohn}H. Spohn, \textit{Large Scale Dynamics of Interacting Particles} (Springer, Heidelberg, 1991).

\bibitem{derrida92}B. Derrida, E. Domany, and D. Mukamel, An exact solution of a one-dimensional asymmetric exclusion model with open boundaries, \textit{J. Stat. Phys.} \textbf{69}, 667 (1992).

\bibitem{schutz}G. M. Sch\"{u}tz, Exactly solvable models for many-body systems far from equilibrium, in \textit{Phase Transitions and Critical Phenomena}, Vol.~19, edited by C. Domb and J. L. Lebowitz (Academic Press, San Diego, 2001), pp.~1--251.

\bibitem{derrida}B. Derrida and J. L. Lebowitz, Exact large deviation function in the asymmetric exclusion process, \textit{Phys. Rev. Lett.} \textbf{80}, 209 (1998); T. Bodineau and B. Derrida, Current fluctuations in nonequilibrium diffusive systems: An additivity principle, \textit{Phys. Rev. Lett.} \textbf{92}, 180601 (2004).

\bibitem{mallick}M. Gorissen, A. Lazarescu, K. Mallick, and C. Vanderzande, Exact current statistics of the asymmetric simple exclusion process with open boundaries, \textit{Phys. Rev. Lett.} \textbf{109}, 170601 (2012); T. Imamura, K. Mallick, and T. Sasamoto, Large deviations of a tracer in the symmetric exclusion process, \textit{Phys. Rev. Lett.} \textbf{118}, 160601 (2017).

\bibitem{krauth}S. C. Kapfer and W. Krauth, Irreversible local Markov chains with rapid convergence towards equilibrium, \textit{Phys. Rev. Lett.} \textbf{119}, 240603 (2017).

\bibitem{pipkin}C. T. MacDonald, J. H. Gibbs, and A. C. Pipkin, Kinetics of biopolymerization on nucleic acid templates, \textit{Biopolymers} \textbf{6}, 1 (1968); C. T. MacDonald and J. H. Gibbs, Concerning the kinetics of polypeptide synthesis of polyribosomes, \textit{Biopolymers} \textbf{7}, 707 (1969).

\bibitem{zia}L. B. Shaw, R. K. Zia, and K. H. Lee, Totally asymmetric exclusion process with extended objects: A model for protein synthesis, \textit{Phys. Rev. E} \textbf{68}, 021910 (2003); T. Chou, K. Mallick, and R. K. P. Zia, Non-equilibrium statistical mechanics: From a paradigmatic model to biological transport, \textit{Rep. Prog. Phys.} \textbf{74}, 116601 (2011).

\bibitem{parmeggiani}I. Neri, N. Kern, and A. Parmeggiani, Totally asymmetric simple exclusion process on networks, \textit{Phys. Rev. Lett.} \textbf{107}, 068702 (2011); I. Neri, N. Kern, and A. Parmeggiani, Modeling cytoskeletal traffic: An interplay between passive diffusion and active transport, \textit{Phys. Rev. Lett.} \textbf{110}, 098102 (2013).

\bibitem{arita}C. Arita, Queueing process with excluded-volume effect, \textit {Phys. Rev. E} \textbf{80} (2009) 051119; C. Arita and D. Yanagisawa, Exclusive queueing process with discrete time, \textit{J. Stat. Phys.} \textbf{141} (2010) 829.

\bibitem{rajewsky}N. Rajewsky, L. Santen, A. Schadschneider, and M. Schreckenberg, The asymmetric exclusion process: Comparison of update procedures, \textit{J. Stat. Phys.} \textbf{92}, 151 (1998).

\bibitem{chowdhury}D. Chowdhury, L. Santen, and A. Schadschneider, Statistical physics of vehicular traffic and some related systems, \textit{Phys. Rep.} \textbf{329}, 199 (2000).

\bibitem{traffic}M. Schreckenberg, A. Schadschneider, K. Nagel, and N. Ito, Discrete stochastic models for traffic flow, \textit {Phys. Rev. E} \textbf{51}, 2939 (1995).


\bibitem{p2p}C. Gkantsidis, M. Mihail, and A. Saberi, Random walks in peer-to-peer networks: Algorithms and evaluation, \textit{Perform. Eval.} \textbf{63}, 241 (2006).

\bibitem{adhoc}S. Srinivasa and M. Haenggi, A statistical mechanics-based framework to analyze ad hoc networks with random access, \textit{IEEE Trans. Mob. Comput.} \textbf{11}, 618 (2012).

\bibitem{aldous83}D. Aldous, Random walks on finite groups and rapidly mixing Markov chains, in \textit{S\'{e}minaire de Probabilit\'{e}s (Strasbourg) XVII 1981/1982}, edited by J. Az\'{e}ma and M. Yor (Springer, Berlin, 1983), pp.~243--297.

\bibitem{aldous86}D. Aldous and P. Diaconis, Shuffling cards and stopping times, \textit{Amer. Math. Month.} \textbf{93}, 333 (1986).

\bibitem{aldfill}D. Aldous and J. A. Fill, \textit{Reversible Markov Chains and Random Walks on Graphs}, unfinished monograph, 2002 (recompiled version, 2014). Available at: \url{http://www.stat.berkeley.edu/~aldous/RWG/book.html}.

\bibitem{persi}P. Diaconis, \textit{Group Representations in Probability and Statistics} (Institute of Mathematical Statistics, Hayward, CA, 1988).


\bibitem{cecche}T. Ceccherini-Silberstein, F. Scarabotti, and F. Tolli, \textit{Harmonic Analysis on Finite Groups} (Cambridge University Press, Cambridge, 2008).


\bibitem{mixing}D. A. Levin and Y. Peres, \textit{Markov Chains and Mixing Times}, 2nd ed. (AMS, Providence, 2017).

\bibitem{survey}L. Lov\'{a}sz, Random walks on graphs: A survey, in \textit{Combinatorics, Paul Erd\H{o}s is Eighty (Keszthely, 1993)}, Vol.~2, edited by D. Mikl\'{o}s, V.~T. S\'{o}s, and T. Sz\H{o}nyi (J\'{a}nos Bolyai Math. Society, Budapest, 1996), pp.~353--398.

\bibitem{saloff}L. Saloff-Coste, Random walks on finite groups, in \textit{Probability on Discrete Structures}, edited by H. Kesten (Springer, Berlin, 2004), pp.~263--346.

\bibitem{kessel}A. Ke{\ss}el, H. Kl\"{u}pfel, J. Wahle, and M. Schreckenberg, Microscopic simulation of pedestrian crowd motion, in \textit{Pedestrian and Evacuation Dynamics 2001}, edited by M. Schreckenberg and S. D. Sharma (Springer, Berlin, 2002), pp.~193--202.

\bibitem{wolki}M. W\"{o}lki, A. Schadschneider, and M. Schreckenberg, Asymmetric exclusion processes with shuffled dynamics, \textit{J. Phys. A: Math. Gen.} \textbf{39}, 33 (2006).

\bibitem{rolland}C. Appert-Rolland, J. Cividini, and H. Hilhorst, Frozen shuffle update for an asymmetric exclusion process on a ring, \textit{J. Stat. Mech.} \textbf{2011}, P07009 (2011).


\bibitem{jungreis}S. Handjani and D. Jungreis, Rate of convergence for shuffling cards by transpositions, \textit{J. Theor. Probab.} \textbf{9}, 983 (1996).

\bibitem{caputo}P. Caputo, T. M. Liggett, and T. Richthammer, Proof of Aldous' spectral gap conjecture, \textit{J. Amer. Math. Soc.} \textbf{23}, 831 (2010).

\bibitem{cesi}F. Cesi, On the eigenvalues of Cayley graphs on the symmetric group generated by a complete multipartite set of transpositions, \textit{J. Algeb. Combin.} \textbf{32}, 155 (2010); A. B. Dieker, Interlacings for random walks on weighted graphs and the interchange process, \textit{SIAM J. Discrete Math.} \textbf{24}, 191 (2010).

\bibitem{kozma}G. Alon and G. Kozma, Ordering the representations of $S_n$ using the interchange process, \textit{Canad. Math. Bull.} \textbf{56}, 13 (2013); G. Alon and G. Kozma, The probability of long cycles in interchange processes, \textit{Duke Math. J.} \textbf{162}, 1567 (2013); N. Berestycki and G. Kozma, Cycle structure of the interchange process and representation theory, \href{https://arxiv.org/abs/1205.4753}{arXiv:1205.4753~[math.PR]}.

\bibitem{nachtergaele}P. Caputo and F. Martinelli, Relaxation time of anisotropic simple exclusion processes and quantum Heisenberg models, \textit{Ann. Appl. Probab.} \textbf{13}, 691 (2003); B. Nachtergaele, W. Spitzer, and S. Starr, Ferromagnetic ordering of energy levels, \textit{J. Stat. Phys.} \textbf{116}, 719 (2004); B. Nachtergaele, W. Spitzer, and S. Starr, Asymptotic ferromagnetic ordering of energy levels for the Heisenberg model on large boxes, \href{https://arxiv.org/abs/1509.00907}{arXiv:1509.00907~[math-ph]}.

\bibitem{cyclestar}B. Morris, Spectral gap for the interchange process in a box, \textit{Electron. Commun. Probab.} \textbf{13}, 311 (2008); M. Conomos and S. Starr, Asymptotics of the spectral gap for the interchange process on large hypercubes, \textit{J. Stat. Mech.} \textbf{2011}, P10018 (2011).

\bibitem{ssepg}J. R. G. Mendon\c{c}a, Exact eigenspectrum of the symmetric simple exclusion process on the complete, complete bipartite and related graphs, \textit{J. Phys. A: Math. Theor.} \textbf{46}, 295001 (2013).

\bibitem{shah81}P. Diaconis and M. Shahshahani, Generating a random permutation with random transpositions, \textit{Z. Wahrsch. verw. Geb.} \textbf{57}, 159 (1981); D. Bayer and P. Diaconis, Trailing the dovetail shuffle to its lair, \textit{Ann. Probab.} \textbf{2}, 294 (1992).

\bibitem{brualdi}R. A. Brualdi and H. J. Ryser, \textit{Combinatorial Matrix Theory} (Cambridge University Press, Cambridge, 1991).

\bibitem{liuchen98}J. S. Liu and R. Chen, Sequential Monte Carlo methods for dynamic systems, \textit{J. Am. Stat. Assoc.} \textbf{93}, 1032 (1998).

\bibitem{pd-rlg-sph}P. Diaconis, R. L. Graham, and S. P. Holmes, Statistical problems involving permutations with restricted positions, in \textit{State of the Art in Probability and Statistics: Festschrift for Willem R. van Zwet}, edited by M. de Gunst, C. Klaassen, and A. Van der Vaart (Institute of Mathematical Statistics, Beachwood, 2001), pp.~195--222.

\bibitem{cdhl}Y. Chen, P. Diaconis, S. P. Holmes, and J. S. Liu, Sequential Monte Carlo methods for statistical analysis of tables, \textit{J. Am. Stat. Assoc.} \textbf{100}, 109 (2005).


\bibitem{js89}M. Jerrum and A. Sinclair, Approximating the permanent, \textit{SIAM J. Comput.} \textbf{18}, 1149 (1989).

\bibitem{rasmussen}L. E. Rasmussen, Approximating the permanent: A simple approach, \textit{Random Struct. Algor.} \textbf{5}, 349 (1994).

\bibitem{kuznetsov}N. Y. Kuznetsov, Computing the permanent by importance sampling method, \textit{Cybern. Syst. Anal.} \textbf{32}, 749 (1996).

\bibitem{smith}P. Smith and B. Dawkins, Estimating the permanent by importance sampling from a finite population, \textit{J. Stat. Comput. Simul.} \textbf{70}, 197 (2001).

\bibitem{jsv04}M. Jerrum, A. Sinclair, and E. Vigoda, A polynomial-time approximation algorithm for the permanent of a matrix with nonnegative entries, \textit{J. ACM} \textbf{51}, 671 (2004).

\bibitem{sinclair}A. Sinclair, \textit{Algorithms for Random Generation and Counting: A Markov Chain Approach} (Birkh\"{a}user, Boston, 1993).


\bibitem{belachung}B. Bollob\'{a}s and F. R. K. Chung, \textit{SIAM J. Discrete Math.} \textbf{1}, 328 (1988).





\end{thebibliography}
\end{document}